\documentclass[twoside,twocolumn,9pt]{article}
\usepackage{extsizes}
\usepackage[super,sort&compress,comma]{natbib}
\usepackage[version=3]{mhchem}
\usepackage[left=1.5cm, right=1.5cm, top=1.785cm, bottom=2.0cm]{geometry}
\usepackage{balance}
\usepackage{times,mathptmx}
\usepackage{sectsty}
\usepackage{graphicx}
\usepackage{lastpage}
\usepackage[format=plain,justification=justified,singlelinecheck=false,font={stretch=1.125,small,sf},labelfont=bf,labelsep=space]{caption}
\usepackage{float}
\usepackage{flushend}
\usepackage{fancyhdr}
\usepackage{fnpos}
\usepackage[english]{babel}
\usepackage{array}
\usepackage{droidsans}
\usepackage{charter}
\usepackage[T1]{fontenc}
\usepackage[usenames,dvipsnames]{xcolor}
\usepackage{setspace}
\usepackage[compact]{titlesec}
\def\bea{\begin{eqnarray}}
\def\eea{\end{eqnarray}}
\def\ba{\begin{eqnarray}}
\def\ea{\end{eqnarray}}
\def\beq{\begin{eqnarray}}
\def\eeq{\end{eqnarray}}
\def\be{\begin{equation}}
\def\ee{\end{equation}}
\def\lbl{\label}
\def\nn{\nonumber \\}

\def\bm{\begin{math}}
\def\me{\end{math}}

\def\lb{\left (}
\def\rb{\right )}

\usepackage{epstopdf}%This line makes .pdf figures into .pdf - please comment out if not required.

\definecolor{cream}{RGB}{222,217,201}

\begin{document}

\pagestyle{fancy}
\thispagestyle{plain}
\fancypagestyle{plain}{

%%%HEADER%%%
%\fancyhead[C]{\includegraphics[width=18.5cm]{head_foot/header_bar}}
%\fancyhead[L]{\hspace{0cm}\vspace{1.5cm}\includegraphics[height=30pt]{head_foot/journal_name}}
%\fancyhead[R]{\hspace{0cm}\vspace{1.7cm}\includegraphics[height=55pt]{head_foot/RSC_LOGO_CMYK}}
\renewcommand{\headrulewidth}{0pt}
}
%%%END OF HEADER%%%

%%%PAGE SETUP - Please do not change any commands within this section%%%
\makeFNbottom
\makeatletter
\renewcommand\LARGE{\@setfontsize\LARGE{15pt}{17}}
\renewcommand\Large{\@setfontsize\Large{12pt}{14}}
\renewcommand\large{\@setfontsize\large{10pt}{12}}
\renewcommand\footnotesize{\@setfontsize\footnotesize{7pt}{10}}
\makeatother

\makeatletter
\renewcommand\@biblabel[1]{#1}
\renewcommand\@makefntext[1]%
{\noindent\makebox[0pt][r]{\@thefnmark\,}#1}
\makeatother
\renewcommand{\figurename}{\small{Fig.}~}
\sectionfont{\sffamily\Large}
\subsectionfont{\normalsize}
\subsubsectionfont{\bf}
\setstretch{1.125} %In particular, please do not alter this line.
\setlength{\skip\footins}{0.8cm}
\setlength{\footnotesep}{0.25cm}
\setlength{\jot}{10pt}
\titlespacing*{\section}{0pt}{4pt}{4pt}
\titlespacing*{\subsection}{0pt}{15pt}{1pt}
%%%END OF PAGE SETUP%%%

%%%FOOTER%%%
%\fancyfoot{}
%\fancyfoot[LO,RE]{\vspace{-7.1pt}\includegraphics[height=9pt]{head_foot/LF}}
%\fancyfoot[CO]{\vspace{-7.1pt}\hspace{13.2cm}\includegraphics{head_foot/RF}}
%\fancyfoot[CE]{\vspace{-7.2pt}\hspace{-14.2cm}\includegraphics{head_foot/RF}}
%\fancyfoot[RO]{\footnotesize{\sffamily{1--\pageref{LastPage} ~\textbar  \hspace{2pt}\thepage}}}
%\fancyfoot[LE]{\footnotesize{\sffamily{\thepage~\textbar\hspace{3.45cm} 1--\pageref{LastPage}}}}
%\fancyhead{}
\renewcommand{\headrulewidth}{0pt}
\renewcommand{\footrulewidth}{0pt}
\setlength{\arrayrulewidth}{1pt}
\setlength{\columnsep}{6.5mm}
\setlength\bibsep{1pt}
%%%END OF FOOTER%%%

%%%FIGURE SETUP - please do not change any commands within this section%%%
\makeatletter
\newlength{\figrulesep}
\setlength{\figrulesep}{0.5\textfloatsep}

\newcommand{\topfigrule}{\vspace*{-1pt}%
\noindent{\color{cream}\rule[-\figrulesep]{\columnwidth}{1.5pt}} }

\newcommand{\botfigrule}{\vspace*{-2pt}%
\noindent{\color{cream}\rule[\figrulesep]{\columnwidth}{1.5pt}} }

\newcommand{\dblfigrule}{\vspace*{-1pt}%
\noindent{\color{cream}\rule[-\figrulesep]{\textwidth}{1.5pt}} }

\makeatother
%%%END OF FIGURE SETUP%%%

%%%TITLE, AUTHORS AND ABSTRACT%%%
\twocolumn[
  \begin{@twocolumnfalse}
%\vspace{3cm}
\sffamily
\noindent\LARGE{\textbf{Regimes of strong   electrostatic collapse of a highly charged polyelectrolyte in a poor solvent.}} \\%Article title goes here instead of the text "This is the title"
 \vspace{0.3cm} \\

\noindent\large{Anvy Moly Tom,$^{\ast}$\textit{$^{a}$} Satyavani Vemparala,\textit{$^{a}$}
 R Rajesh,\textit{$^{a}$} and Nikolai Brilliantov\textit{$^{b}$} }\\%Author names go here instead of "Full name", etc.

\noindent\normalsize{We perform extensive
molecular dynamics simulations of a highly charged flexible polyelectrolyte (PE) chain in a poor solvent for
the case when the chain is in a collapsed state and the electrostatic interactions, characterized by the
reduced Bjerrum length $\ell_B$, are strong. We detect the existence of several sub-regimes, $R_g \sim
\ell_B^{-\gamma}$, in the dependence of the gyration radius of the chain $R_g$ on $\ell_B$. In contrast to a
good solvent,  the exponent $\gamma$ for a poor solvent crucially depends  on the size and valency of
counterions. To explain the different sub-regimes we generalize the existing counterion fluctuation theory by
a more complete account of the volume interactions in the free energy of the chain. These include
interactions between the chain monomers, between monomers and counterions and the counterions themselves. We
also demonstrate that the presence of the condensed counterions can modify the effective attraction among the
chain monomers and impact the sign of the second virial coefficient.
} \\%The abstrast goes here instead of the text "The abstract should be..."
\end{@twocolumnfalse} 
\vspace{0.6cm}

  ]
%%%END OF TITLE, AUTHORS AND ABSTRACT%%%

%%%FONT SETUP - please do not change any commands within this section
\renewcommand*\rmdefault{bch}\normalfont\upshape
\rmfamily
\section*{}
\vspace{-1cm}

%%%FOOTNOTES%%%

\footnotetext{\textit{$^{a}$~The Institute of Mathematical Sciences, C.I.T. Campus,Taramani, Chennai 600113, India }}
\footnotetext{\textit{$^{b}$~Department of Mathematics, University of Leicester, Leicester LE1 7RH,United Kingdom }}

%Please use \dag to cite the ESI in the main text of the article.
%If you article does not have ESI please remove the the \dag symbol from the title and the footnotetext below.
\footnotetext{\dag~Electronic Supplementary Information (ESI) available: [details of any supplementary information available should be included here]. See DOI: 10.1039/b000000x/}
%additional addresses can be cited as above using the lower-case letters, c, d, e... If all authors are from the same address, no letter is required

\footnotetext{\ddag~Additional footnotes to the title and authors can be included \textit{e.g.}\ `Present address:' or `These authors contributed equally to this work' as above using the symbols: \ddag, \textsection, and \P. Please place the appropriate symbol next to the author's name and include a \texttt{\textbackslash footnotetext} entry in the the correct place in the list.}

%%%END OF FOOTNOTES%%%

%%%MAIN TEXT%%%%
\section{Introduction.}
Charged polymers in solution, or polyelectrolytes (PEs), are ubiquitous in nature and play a significant role
in our everyday life. Common examples of PEs include biologically important molecules  like DNA, RNA and
proteins~\cite{PhysRevE.71.061928,BIP:BIP360311305,BLOOMFIELD1996334}. PEs also find application in
industries such as chemical
\cite{zhao2011polyelectrolyte,ANIE:ANIE200805456,renault2009chitosan,fang2002novel}, pharmaceutical
\cite{B808262F,rew-lan,Shu2010210,Anandhakumar2011342}, food\cite{ANIE:ANIE2201,shahidi1999food} etc. The
mechanical and chemical properties of a PE depend on its conformational state, which could vary from being
linear and extended to compact and collapsed. The conformational state is determined essentially by three
properties: the strength of electrostatic interactions in the system,  entropy of the counterions and quality
of the solvent. Determining the precise role of these characteristic in the conformational state of a simple
model PE is fundamental to understand the physics of more realistic PE systems.

The strength of the electrostatic interactions in the system  depends on the charge density of the PE chain
and is quantified by the reduced Bjerrum length $\ell_B$
\be
\lbl{eq:BL}
\ell_B = \frac{e^2}{\epsilon k_B T
a}=\frac{\beta e^2}{\epsilon a}, \ee
where $\epsilon$ is the dielectric permittivity of the solvent, $k_B$ is the Boltzmann constant, $T$ is
temperature, $\beta =(k_BT)^{-1}$, and $a$ is the distance between neighbouring charged monomers of the PE
chain. $\ell_B$ quantifies the ratio of the electrostatic interaction energy between the neighbouring charged
monomers and thermal energy. The larger the value of $\ell_B$, the stronger the electrostatic interactions in
the system.

For small $\ell_B$ a  PE behaves like a neutral polymer and the counterions are dispersed away from the PE to
occupy all accessible volume, resulting in a state with high entropy. With increasing $\ell_B$, the
electrostatic interaction energy becomes comparable to the thermal energy and counterions begin to condense
onto the PE, renormalizing its charge density~\cite{manning69}. The condensed counterions being in a close
vicinity of the PE imply a lower entropy of the system. Still, if $\ell_B$ is not large, a PE conformation is
dictated by the solvent quality.

The solvent quality, in turn,  is determined by the relative strength of the attractive interactions between
chain monomers and between monomers and solvent particles. In a good or theta-solvent these attractive interactions are
stronger for monomer-solvent particle pairs, while in a poor solvent monomer-monomer attractive interactions
dominate~\cite{khokhlov1994}.

Several experiments and simulations have shown that at large enough $\ell_B$, like-charged PE chains undergo
a transition from  extended to collapsed conformations regardless of the solvent quality
~\cite{Kremer1993,Kremer1995,winkler98,brilliantov98,BoJon99,Holm,anoop11,Dobrynin2005,Chertovich2016}. This
counterintuitive transition is driven by the condensation of counterions onto the chain, reducing the
effective charge density. The nature of the effective attractive interactions driving the transition is not
well understood and there are competing theories explaining their origin (see below). For the collapsed
state, these theories predict that the gyration radius, $R_g$, of a PE has the  scaling  form $R_g \sim
N^{1/3} \ell_B^{-\gamma}$,  where  $N$ is the length of the PE, and the exponent $\gamma$ can potentially
depend on system parameters. Presently there exist three different theoretical approaches to explain this
electrostatics-driven collapse in
PEs~\cite{brilliantov98,pincus98,Kardar1999,delaCruz2000,Cherstvy2010,muthukumar2004,arti14}. These theories
differ from each other in the way the effective electrostatic interactions are modelled and lead to different
predictions for the exponent $\gamma$. In the first approach it is argued that the free energy of a PE in a
collapsed state  corresponds to that of an amorphous ionic solid; this theory predicts a collapsed
conformation for the case of multivalent counterions, with no dependence of $R_g$ on $\ell_B$
($\gamma=0$)~\cite{delaCruz2000}. In the second approach it is hypothesized that the freely rotating
fluctuating dipoles formed between PE monomers and condensed counterions give rise to an effective attractive
interaction between segments of the chain, which causes a PE collapse both in good and poor
solvents~\cite{pincus98,Cherstvy2010,muthukumar2004,arti14}. This theory predicts that $R_g$ of a PE in a
collapsed state scales with Bjerrum length as  $R_g \sim  N^{1/3}\left| \ell_B^2 - c B_2 \right|^{-1/3} $,
where $B_2$ is the second virial coefficient, and $c$ is a dimensional constant that depends on the details
of the system~\cite{pincus98,muthukumar2004,arti14}. Hence for both
good~\cite{pincus98,muthukumar2004,arti14} and poor~\cite{pincus98,arti14} solvents, $\gamma$ is equal to
 $2/3$. In the third approach, known as counterion-fluctuation theory,  it is argued that density fluctuations
of condensed counterions inside a chain globule cause a negative pressure, which drives the PE
collapse~\cite{brilliantov98}. This theory predicts $\gamma=1/2$.

In a recent molecular dynamics (MD) study of a flexible PE in a good solvent we showed that a
collapsed PE conformation demonstrates at least two different sub-regimes, which we call as a weak
electrostatic collapse, with $\gamma =1/2$,  and a strong electrostatic collapse with $\gamma=1/5$. The
exponent $\gamma$ in both regimes was  found to be independent of the valency of counterions and PE chain
length~\cite{CollGood}. All inter-particle interactions, other than electrostatic, were repulsive in these
systems.  The counterion-fluctuation theory, originally developed for good solvent with a single exponent
$\gamma=1/2$, has been   generalized by us~\cite{CollGood} through the inclusion of higher order virial
coefficients  to reproduce both the weak ($\gamma=1/2$) and strong ($\gamma=1/5$) collapse regimes seen in
our MD simulations.

It is more challenging however to study a collapsed state of a flexible PE in a poor solvent ~\cite{Chang2003, Micka1999, Micka2000, Thirumalai2001, Chang2006, limbach03, loh2008} since, unlike in
a good solvent,  there exist additional attractive interactions between monomers which compete with the
repulsive part of electrostatic interactions. The  valency of counterions, which played no role in determining the
exponent $\gamma$ for a good solvent becomes significant for a poor solvent. Indeed, it dictates  the number
of counterions condensed inside the collapsed globule; the presence of the counterions modifies the overall
interaction energy between the monomers due to the excluded volume interactions between all species and hence
influences the exponent $\gamma$.   In the present study, we report MD simulations for the collapsed phase of
a strongly charged flexible PE in a poor solvent and find several novel collapse sub-regimes. The observed in
MD simulations  conformational behavior of a PE in a poor solvent can be explained by extending the modified
counterion-fluctuation theory~\cite{brilliantov98,CollGood} with a more complete account of the volume
interactions between all species in the system: the monomer-monomer, monomer-counterion and
counterion-counterion interactions. We show that the new theory can uniformly describe the MD results for
both good and poor solvents.

The rest of the paper is organized as follows. In Sec.~\ref{sec:MD}, we define the model and give details of
the MD simulations.  In Sec.~\ref{sec:results}, we present our theory of the electrostatic collapse in a poor
solvent and compare the predictions with data from MD simulations. A summary and discussion of our results
are given in Sec.~\ref{sec:discussion}.

\section{Model and Methods \label{sec:MD}}

We model a flexible PE chain as $N$ monomers of charge $e$  connected by harmonic springs of energy \be
U_{bond}(r)=\frac{1}{2} k(r-a)^2, \ee where $k$ is the spring constant, $a$ is the equilibrium bond length,
and $r$ is  the instantaneous distance between the bonded monomers. The PE chain and $N_c=N/Z$ neutralizing
counterions with a valency $Z$ are  placed in a box of linear size $L$. Pairs of all non-bonded particles
(counterions and monomers) separated by a distance $r_{ij}$ interact through the volume (or van der Waals)
interactions. Here we model these interactions by the Lennard Jones (LJ) $6-12$ potential with a cutoff of
$r_c$:
\begin{equation}
\lbl{eq:LJpot}
U_{LJ}(r_{ij})=4\epsilon_{LJ} \left[\left(\sigma/r_{ij} \right)^{12}-\left(\sigma/r_{ij} \right)^6 \right], ~ r_{ij} \leq r_c .
\end{equation}
The values of $\epsilon_{LJ}$ and $r_c$ are varied to model solvents of different quality  (see below). The
electrostatic energy between charges $q_i$ and $q_j$ separated by $r_{ij}$ is given by
\begin{equation}
U_{c}(r_{ij})= \frac{q_iq_j}{\epsilon r_{ij}}.
\label{eq.1}
\end{equation}
In the simulations, we use $a=1.12 \sigma$, $k=500.0 k_B T/\sigma^2$, $L=370 \sigma$, N=204.  All energies
are measured in units of  $k_B T$, and we maintain temperature at $1$  through a  Nos\'{e}-Hoover thermostat.
All distances are measured in terms of $\sigma$ which we set to 1. The long-range Coulomb interactions are
evaluated using particle-particle/particle-mesh (PPPM) technique.

The equations of motion are integrated in time using molecular  dynamics simulation package
LAMMPS~\cite{plimpton1995fast} with a time step of  $0.001$. All the  systems are equilibrated for
$5\times10^6$ timesteps and the data presented in this paper are averaged over $5\times10^6$ timesteps of
production runs.

We model a variety of poor solvent conditions by choosing the  following combinations of the LJ energy
parameters for  $\epsilon _{LJ}$ and cutoff distance $r_c$ for the monomer-monomer interactions: (i)
$\epsilon _{LJ}=1$ and  $r_c=2.5$,  (ii) $\epsilon _{LJ}=1.5$ and  $r_c=2.5$, and (iii) $\epsilon _{LJ}=2$
and  $r_c=2.5$.  For all other volume interactions among counterions and between monomers and counterions,
the LJ interactions are repulsive, $\epsilon _{LJ}=1.0$ and $r_c=1.0$. These conditions are chosen in such a
 way, that when the charge on the monomers is zero, a PE chain adopts a collapsed conformation, mimicking poor
solvent conditions. We also performed additional simulations in which the counterion size was varied. We note
that all simulations reported in this paper have been performed for the values of $\ell_B$, where the
equilibrium configuration of a PE is a collapsed state with $R_g \sim N^{1/3}$.

\section{Results \label{sec:results}}

To develop a generalized theory for electrostatic-driven collapse  of a PE in a poor solvent, we start with
the modified counterion-fluctuation theory for a good solvent and retain the electrostatic term.  As
mentioned in the Introduction, the modified counterion fluctuation theory is able to explain the observed in
MD simulations collapsed sub-regimes (with correct exponent $\gamma$)  for a flexible PE  in a good
solvent~\cite{CollGood}. However, in the case of a poor solvent, we anticipate that  the effective attraction
between chain monomers, supplemented by attractive van der Waals forces, would cause even stronger collapse
of a chain compared to that in a good solvent. In the following subsections, we generalize the theory of
Ref.~\cite{CollGood} making a more complete account of the volume interactions and show that the
counterion-fluctuation theory is applicable regardless of a solvent quality.

\subsection{\label{sec:theory}Free energy of a  PE system}

To find the equilibrium gyration radius $R_g$ we compute  the conditional free energy of the system $F(R_g)$
and minimize it with respect to $R_g$. We write the free energy of the system  as a sum of its components as
follows: \be \lbl{eq:FRg}
 F(R_g) =F_{\rm id. ch}(R_g)+F_{\rm en}(R_g)+ F_{\rm el}(R_g)+F_{\rm vol} (R_g),
\ee where $F_{\rm id.ch}$,  $F_{\rm en}$, $F_{\rm el}$, and $F_{\rm vol}$ account for the free energy of an
ideal chain, entropy of the counterions, the electrostatic interactions between the charged particles, and
the volume interactions between all the species respectively.

$F_{\rm id.ch}(R_g)$, the part of the free energy corresponding to the ideal chain
reads~\cite{GrosbergKuznetsov_MM92,brilliantov98,khokhlov1994},
\begin{equation}
\label{eq:Fn}
\beta F_{\rm id.ch} \simeq  \frac94 \left(\alpha^2 + \alpha^{-2} \right),
\end{equation}
where $\alpha = R_g/R_{\rm g. id}$ is the expansion factor, with $R_{\rm g.id}$   being the gyration radius
of the ideal chain, $R^2_{\rm g.id}=Na^2/6$.

The second part of the system free energy, $F_{\rm en}(R_g)$ accounting  for the  entropy of  the counterions
is proportional to the logarithm of the volume available for the counterions. It may be shown
that~\cite{brilliantov98,CollGood},
\begin{equation}
\label{eq:en}
\frac{\beta F_{\rm en}}{N}=-\frac{3}{Z}(1-\tilde{\rho}) \ln \left(\frac{R_0}{a}\right),
\end{equation}
where $\tilde{\rho}=\rho_{\rm c. in}/\rho_0$  with $\rho_{\rm c. in}$  being the number density of
counterions within the globule with the gyration volume $V_{\rm g} = 4\pi R_g^3/3$ and $\rho_0=N_c/V_{\rm
g}=N/(ZV_{\rm g})$ is the maximal counterion number density, when their condensation is complete. The value
of $R_0$ refers to the volume $4\pi R_0^3/3$ per chain in the solution; it corresponds  to $L$ in  our  MD
simulations.

The third part of the system free energy, $F_{\rm el}(R_g)$,  accounting for the electrostatic interactions
among monomers and  counterions is given by the  counterion fluctuation theory~\cite{brilliantov98} as:
\begin{equation}
\label{eq:el}
\frac{\beta F_{\rm el}}{N}=\frac{3\sqrt{6} \ell_B N^{1/2}(1-\tilde{\rho})^2}{5\alpha} \left(1-\frac{2R_g}{3 R_0} \right) -
\frac32 \left(\frac{2}{\pi^2} \right)^{1/3} \frac{\ell_B \sqrt{6} Z^{2/3} \tilde{\rho}^{4/3}}{N^{1/6} \alpha}.
\end{equation}
The first term in the right hand side of  \eqref{eq:el} gives  the mean-field result for the electrostatic
interactions in the system. The second term describes the contribution to the free energy from the correlated
fluctuations of the charge density and is beyond the Poisson-Boltzmann approximation~\cite{brilliantov98}.
Both Eqs.~\eqref{eq:en} and \eqref{eq:el} refer to dilute salt-free PE solutions of long chains, $R_0 \gg
R_g$ and $N \gg 1$~\cite{brilliantov98}.

Finally,  the free energy $F_{\rm vol} (R_g)$  accounting  for the volume (LJ) interactions between monomers
and counterions may be written as
\be
\lbl{eq:Fvol} F_{\rm vol} (R_g)  = F_{\rm vol\, m.m.}+F_{\rm vol\,
c.m.}+F_{\rm vol\, c.c.},
\ee
where $F_{\rm vol\, m.m.}$, $F_{\rm vol\, c.m.}$, and $F_{\rm vol\, c.c.}$ are the free energy terms for the
volume interactions between monomers, counterions and monomers, and between counterions respectively.

The monomer packing fractions, computed for the collapsed phase of the flexible PE in poor solvent considered here, vary from 0.1 to 0.25 
across the range of electrostatic strengths considered here. Since the measured packing fractions are not too large~\cite{khokhlov1994}, 
we write the free energy as a virial expansion. Keeping up to the third virial term, we obtain for $F_{\rm vol\, m.m.}$: 
\bea
\lbl{eq:Fvol_mm1} \beta F_{\rm vol\,
m.m.} &=& \lb B_2 \rho_m^2+B_3 \rho_m^3 \rb \, V_{\rm g}
= \lb B_2 \left(\frac{N}{V_{\rm g}} \right)^2  +B_3 \left(\frac{N}{V_{\rm g}} \right)^3 \rb \, V_{\rm g} \nonumber \\
&=&b^{-1} \frac{B_2}{\alpha^3 N^{1/2}} +b^{-2} \frac{B_3}{\alpha^6}\,,
\eea
where $B_2$ and $B_3$ are the second and third virial coefficients for monomer-monomer interactions, $\rho_m = N/V_{\rm g}$ is the
average density of monomers inside the gyration volume and $b=\lb 2 \pi a^3 /9 \sqrt{6} \rb$. For a collapsed
state addressed here almost all counterions are located within the gyration volume of the chain. Then the
average counterion density inside the gyration volume will be $\rho_{\rm c.in} \simeq N_c/V_{\rm
g}=N/(ZV_{\rm g})$  and we neglect the counterion density outside this volume. The  free energy of the volume
interactions of the counterions will have the same form as in Eq.~(\ref{eq:Fvol_mm1}), but with the virial
coefficients $C_k$ for the counterion-counterion interactions, divided by $Z^k$, where $Z$ is the valency of
the counterions for the $k$-th virial term. A similar expression follows for  the volume interactions between the chain monomers and
counterions. Combining these expressions for $F_{\rm vol\, m.m.}$, $F_{\rm vol\, c.m.}$ and $F_{\rm vol\,
c.c.}$  and limiting to third virial term, one arrives at the following result for the free energy $F_{\rm vol}$ for the case of complete
counterion condensation:

\bea
\lbl{eq:Fvol_all1} \beta
F_{\rm vol} = \frac{\tilde{B}_2}{\alpha^3 N^{1/2}} + \frac{\tilde{B}_3}{\alpha^6}\,,
\eea
where $\tilde{B}_2$  and $\tilde{B}_3$are the
renormalized second and third virial coefficients respectively that account for all volume  interactions. These coefficients read: \bea
\lbl{eq:Btild1}
\tilde{B}_2 &=&b^{-1} \lb B_2+\frac{2 D_{1,1}}{Z} + \frac{C_2}{Z^2} \rb \\
\tilde{B}_3 &=&b^{-2} \lb B_3+ \frac{3D_{1,2}}{Z} + \frac{3D_{2,1}}{Z^2} + \frac{C_3}{Z^3} \rb, \eea where
$C_k$ is $k$-th virial coefficient for the counterion-counterion  interactions and $D_{l, k-l}$ are the
virial coefficient for monomer-counterion volume interactions of $l$-th order with respect to  the
counterions and $(k-l)$-th order with respect to monomers (see the Appendix for the detail). The value of the
virial coefficients $B_k$, $C_k$ and $D_{l, k-l}$ are determined by the relative strength of the LJ
interactions and the thermal energy $k_B T$. Due to the dominance of repulsive forces in the
monomer-counterion and counterion-counterion volume interactions considered here,  all coefficients $C_k$ and
$D_{l, k-l}$ are expected to be positive, $C_k>0$ and $D_{l, k-l}>0$ for $k\geq 2$ and $1 \leq l \leq k$. We
also assume that the  renormalized third virial coefficient $\tilde{B}_3$ is positive as well.

At the same time the sign of the renormalized second virial coefficient  $\tilde{B}_2$ sensitively depends
not only on its "bare" value $B_2$ (which is negative for poor solvents addressed here), but also on the
counterion valency and the virial coefficients $C_k$ and $D_{l, k-l}$, see Eq.~\eqref{eq:Btild1}.  If these
positive coefficients are large and the valency $Z$ is small,  $\tilde{B}_2$ becomes positive even for
negative $B_2$. Oppositely, for small $C_k$ and $D_{l, k-l}$ and large $Z$ the renormalized second virial
coefficient remains negative,  $\tilde{B}_2<0$. The values of the virial coefficients $C_k$ and $D_{l, k-l}$
are determined by the size of counterions: The larger the counterions, the larger the virial coefficients.
Hence it is expected that small counterions with a high valency imply negative $\tilde{B}_2$, while large
counterions with a low valency imply positive renormalized coefficient, $\tilde{B}_2>0$ (see the Appendix for
the detail). This predictions will be checked in our MD simulations discussed below.

If the packing fraction of species (monomers and counterions) inside the gyration globule increases, the truncated
expansion \eqref{eq:Fvol_mm1} loses its accuracy. One needs to use then an equation of state (EOS) for dense
fluids, which may be the Flory-Huggins or van der Waals EOS with the appropriate parameters describing volume
interactions. One can also use the EOS for Lennard-Jones mixtures, e.g.~\cite{LJmix}. In the case of systems with larger packing fraction,
additional terms in the virial expansion of  \eqref{eq:Fvol_all1} are included, which leads to the general form:
%However, for the sake
%of  \emph{qualitative} analysis reported here we prefer to use the extended virial equation of state, which
%is the generalization of the above expression \eqref{eq:Fvol_all1},
%
\bea
\lbl{eq:Fvol_all} \beta F_{\rm vol} = \sum_{k=2}^{\infty} \frac{N^{(3-k)/2}}{\alpha^{3(k-1)}} \,
\tilde{B}_k ,
\eea
that includes all the virial coefficients,
\be
\lbl{eq:Btild} \tilde{B}_k =b^{1-k} \left(
B_k + \sum_{l=0}^{k} \frac{{\cal C}_l^k}{Z^l}D_{l, k-l} + \frac{C_k}{Z^k} \right),
\ee
where ${\cal C}_l^k = k!/l!(k-l)!$ are  the combinatorial coefficients, see the Appendix for the detail. Note
that any non-singular EOS may be expressed in the form of Eq.~\eqref{eq:Fvol_all}, where the virial
coefficient depend on the particular EOS. Here we use the approach developed for dense gases, where  the
virial coefficients are explicitly expressed in terms of the interaction potential~\cite{Croxton1974}, see
also the Appendix.

For small packing fractions only contribution from the first two terms, as in Eq.~\eqref{eq:Fvol_all1} is
non-negligible. With the increasing density, next order virial terms in \eqref{eq:Fvol_all} start to play a
role. It may happen that in some limited interval of packing fractions one particular term in the expansion
\eqref{eq:Fvol_all} dominates. This will then manifest in a specific behavior of physical quantities for this
interval (see the discussion below).

Combining the different contributions [Eqs.~\eqref{eq:Fn}, \eqref{eq:en}, \eqref{eq:el} and
\eqref{eq:Fvol_all}] the free energy in Eq.~\eqref{eq:FRg} attains the form, \bea &&\frac{\beta
F(R_g)}{N}\simeq \frac{9}{4N} \left(\alpha^2 + \alpha^{-2} \right)
-\frac{3}{Z}(1-\tilde{\rho}) \ln \left(\frac{R_0}{a}\right) \label{eq:free_final}\\
&&+\frac{3\sqrt{6} \ell_B N^{1/2}(1-\tilde{\rho})^2}{5\alpha} \left(1-\frac{2R_g}{3 R_0} \right)
-\frac{\tilde{Z}^2  \ell_B  }{N^{1/6} \alpha}
+\sum_{k=2}^{\infty} \frac{N^{(1-k)/2}}{\alpha^{3(k-1)}} \, \tilde{B}_k, \nonumber
\eea
where $\tilde{Z}^2 =(3/2)(2/\pi^2)^{1/3}\sqrt{6}\,Z^{2/3}$.

 In a collapsed state of a PE, addressed in the present study, almost all counterions are located inside the collapsed globule, regardless of the solvent quality, so that the average density of counterions is close to its maximal density inside the globule, $\rho_0=N_c/V_{\rm g}$. This  suggests the approximation $\tilde{\rho} \approx 1$, allowing us to ignore the  second and third
terms in the right hand side of Eq.~\eqref{eq:free_final}. Moreover, since  $R_g\sim N^{1/3}$ in a collapsed
regime, the expansion factor $\alpha \sim N^{-1/6} \ll 1$. Thus, the term proportional to $\alpha^2$ in
$F_{\rm id.ch}(\alpha)$ can be neglected. It is also straightforward to see that, in this limit, the term
proportional to $\alpha^{-2}$ in  $F_{\rm id.ch}(\alpha)$ is small compared to the volume terms  and may be
dropped as well. With these approximations,  Eq.~\eqref{eq:free_final} for the free energy of a PE in a
collapsed state reduces to the following expression, regardless of the solvent quality:
\begin{equation}
\label{eq:ftot}
\frac{\beta F}{ N} = -\frac{\tilde{Z}^2  \ell_B  }{N^{1/6} \alpha}+\sum_{k=2}^{\infty} \frac{N^{(1-k)/2}}{\alpha^{3(k-1)}} \, \tilde{B}_k .
\end{equation}

\subsection{Scaling of $R_g$ and energies of a PE chain with $\ell_B$ \label{sec32}}

The dependence of gyration radius of a PE chain, $R_g= \alpha R_{\rm g. id}$   on $\ell_B$ may be obtained by
minimizing the free energy, Eq.~\eqref{eq:ftot}, with respect to $\alpha$, which results in, \be \lbl{eq:min}
\frac{\tilde{B}_2}{N^{1/3} }\alpha^{-2} + \frac{2\tilde{B}_3 }{N^{5/6} }\alpha^{-5} + \frac{3\tilde{B}_4
}{N^{4/3} } \alpha^{-8}+ \ldots = \frac{\tilde{Z}^2  \ell_B  }{3}, \ee
where a general term in the left-hand side of the above equation reads $(k-1)\tilde{B}_{k}N^{(4-3k)/6}
\alpha^{4-3k}$. The relative importance of different terms in Eq.~\eqref{eq:min} depends on $N$, the virial
coefficients $\tilde{B}_k$ and the expansion factor $\alpha$. In what follows, we show how the dominance of
different virial terms in Eq.~\eqref{eq:min} gives rise to unique scaling exponents in the dependence of
$R_g$ on $\ell_B$, for a limited interval of $\ell_B$, that is, for a limited interval of species densities
(monomers and counterions) inside the PE globule.

For the case of weak electrostatic collapse, the first term in the left  hand side  of Eq.~\eqref{eq:min}
dominates yielding,
\begin{equation}
\label{eq:lb12}
R_g = \frac{\sqrt{\tilde{B}_2} a N^{1/3}}{\sqrt{2} \tilde{Z}\ell_B^{1/2}},
\end{equation}
that is,  $R_g \sim \ell_B^{-1/2}$. We note that this yields a meaningful  result only if ${\tilde{B}_2}$ is
positive. While this is trivially satisfied in the case of good solvent, for the poor solvent conditions
considered here, the attractive LJ-forces between the monomers will prevail and the bare second virial
coefficient $B_2$ is always negative, $B_2<0$.  However as we show later the opposite condition
${\tilde{B}_2}> 0$ may be satisfied even in the case of poor solvent for certain counterion sizes and
valencies.

When $\ell_B$ increases and  $\alpha$ becomes smaller, the subsequent terms in Eq.~\eqref{eq:min} begin to
dominate over the first term. When the second or third terms on left hand side of Eq.~\eqref{eq:min}
dominate, we obtain respectively the following exponents, $R_g \sim \ell_B^{-1/5}$ and $R_g \sim
\ell_B^{-1/8}$:
\begin{equation}
\label{eq:lb15}
R_g = \frac{ \tilde{B}_3^{1/5} a N^{1/3}}{6^{3/10}\tilde{Z}^{2/5}  \ell_B^{1/5}},
\end{equation}
and
\begin{equation}
\label{eq:lb18} R_g = \frac{ \tilde{B}_4^{1/8} a N^{1/3}}{3^{1/4}\sqrt{2}\tilde{Z}^{1/4}  \ell_B^{1/8}}.
\end{equation}
In a general case, when the $k$-th virial term (with the coefficient $\tilde{B}_k$) dominates in some
interval of $\ell_B$, the gyration radius scales with the reduced Bjerrum length as $R_g \sim
\ell_B^{1/(3k-4)}$. The sequence of the exponents $\gamma$ for the inverse power of $\ell_B$ corresponding to
 $k$-th virial term is shown in Table~\ref{table1}.
\begin{table}
\caption{The values of the exponent $\gamma$ corresponding to the dominance of the  $k^{\mathrm{th}}$ virial
term.} \label{table1}
\begin{center}
\begin{tabular}{|c|c|}
\hline
$k$&$\gamma$\\
\hline
$2$&$1/2$\\\hline
$3$&$1/5$\\\hline
$4$&$1/8$\\\hline
$5$&$1/11$\\\hline
$6$&$1/14$\\\hline
$7$&$1/17$\\\hline
$8$&$1/20$\\\hline
$9$&$1/23$\\\hline
\end{tabular}
\end{center}
\end{table}

We also derive the dependence of the  internal energies, associated with the electrostatic and volume LJ
interactions, on the gyration radius $R_g$. Using  the thermodynamic relation for the internal energy $E
=\partial \beta F/\partial \beta$, we obtain \bea
\beta E_{el}/( N \ell_B)& =& - \tilde{Z}^2 a N^{1/3}/\sqrt{6} R_g \sim  N^{1/3}R_g^{-1},\label{eq:Eel}\\
\frac{\beta E_{LJ}}{N} &=& \frac{\tilde{B}_2^{\prime} N}{R_g^3}+  \frac{\tilde{B}_3^{\prime} N^2}{R_g^6}+
\frac{\tilde{B}_4^{\prime} N^3}{R_g^9}+ \frac{\tilde{B}_5^{\prime} N^4}{R_g^{12}}+ \ldots, \label{eq:ELJ}
\eea where $E_{el}$ and $E_{LJ}$ are the electrostatic and LJ components of the internal energy and the
constants $\tilde{B}_k^{\prime}$ may be expressed in terms of the temperature derivatives of the reduced
virial coefficients $\tilde{B}_k$. These scaling laws can be easily measured in MD simulations.

\subsection{MD results}

\begin{figure}
\centering
\includegraphics[width=\columnwidth]{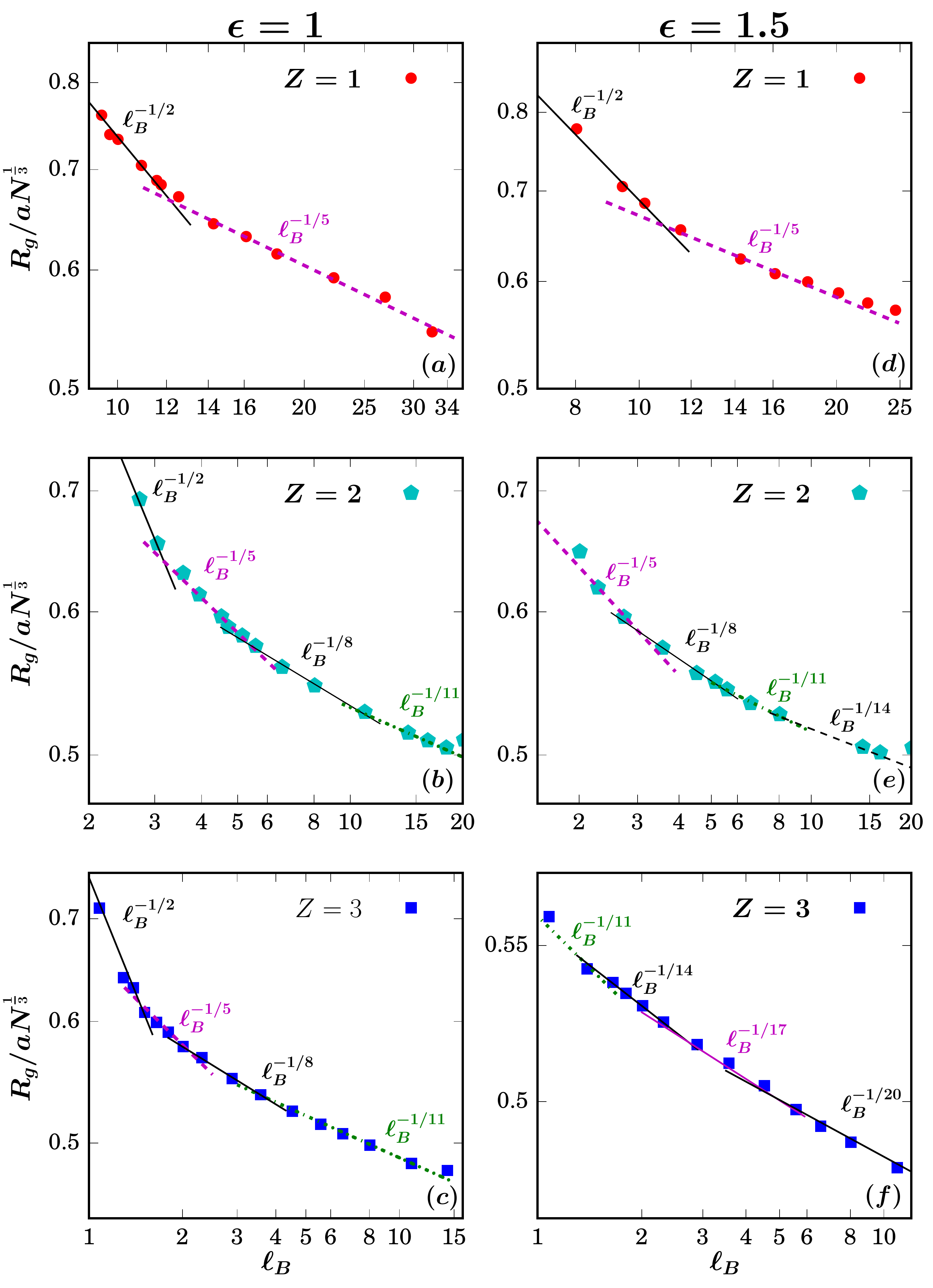}
\caption{Variation of the gyration radius $R_g$  with $\ell_B$ for a PE chain with counterion valency $Z=1$
(a,d); $Z=2$ (b,e) and $Z=3$ (c,f). The data are for poor solvent conditions ($\epsilon_{LJ}=1.0$ and $
\epsilon_{LJ}=1.5$ and $r_c=2.5$  for monomers). The solid straight lines correspond to power laws with
exponents $\gamma$ as predicted by the theory (see Table~\ref{table1}). } \label{fig:RgZ_Poor}
\end{figure}

To understand the effect of solvent quality on the scaling of $R_g$ with $\ell_B$, we have simulated a
flexible PE in various poor solvents. In Fig.~\ref{fig:RgZ_Poor}, we show the variation of $R_g$ for a
collapsed PE with $\ell_B$ for different counterion valency for two poor solvent conditions
($\epsilon_{LJ}=1.0,~1.5$ and $r_c=2.5$ for monomer-monomer interaction). In the case of $\epsilon_{LJ}=1.0$
[see  Fig.~\ref{fig:RgZ_Poor}(a)--(c)], for all three valencies the weak and strong electrostatic collapse
regimes with $\gamma=1/2$ and $\gamma =1/5$ are observed, when the compaction of the chain increases.
However, in the case of divalent and trivalent counterions the additional sub-regimes appear, as $\ell_B$
further increases. These regimes with smaller values of $\gamma$ are observed for different intervals of
$\ell_B$, corresponding to the larger density of the PE globule. When $\epsilon_{LJ}$ is changed to $1.5$
[see Fig.~\ref{fig:RgZ_Poor}(d)--(f)], the weak electrostatic regime with $\gamma=1/2$ persists only for
monovalent counterions and vanishes for divalent and trivalent counterions. At the same time the regimes with
smaller values of the exponent $\gamma$ emerge. Corresponding data for  $\epsilon_{LJ}=2.0$ is given in
Supplementary Information, where it is clearly demonstrated that the weak electrostatic regime with
$\gamma=1/2$ still persists for systems with monovalent counterions. The MD data presented in
Fig.~\ref{fig:RgZ_Poor} are consistent with the theoretically predicted power laws. Both the values of the
exponent $\gamma$ and the sequence of their appearance are in agreement with the theory,  see
Table~\ref{table1}.

The appearance of $\gamma=1/2$ for the poor solvent case shown in Fig.~\ref{fig:RgZ_Poor} is  surprising
since in poor solvent conditions, the second virial coefficient $B_2$, when restricted to monomer-monomer
interactions,  is expected to be negative. Indeed, as we have mentioned above, if the charge of monomers is
zero, the PE always adopts a collapsed state, corresponding to the negative value of
$B_2$~\cite{khokhlov1994}. Then from Eqs.~\eqref{eq:lb12} and \eqref{eq:lb15}, it can be seen that the
largest  possible value for the exponent $\gamma$ should be  $1/5$,  corresponding to the (positive) third
virial term. Hence we conclude that the presence of the counterions inside a collapsed globule leads to the
change of the sign of $B_2$ for a certain range of LJ parameters and valency of counterions. This agrees with
the above analysis, where we stated  that for large counterions with a low valency the effective second
virial coefficient $\tilde{B}_2$ is positive, although the bare coefficient $B_2$ is negative, yielding the
regime with $\gamma=1/2$. Physically, the negative sign of $\tilde{B}_2$ follows from the dominance of
attractive volume interactions. Therefore, large counterions with a low valency (which implies the larger
counterion density inside a globule) keep the chain monomers apart and reduce the effect of attractive volume
interactions between them; this results in the alteration of the sign of $\tilde{B}_2$. For $\tilde{B}_2>0$
the regime with $\gamma =1/2$ is observed. At the same time small counterions of high valency (which implies
lower density of these inside the globule) can not effectively keep the monomers apart, so that their
attractive volume interactions yield a negative $\tilde{B}_2$. For these systems the regime with $\gamma
=1/2$ is absent.

In other words, if the regime with $\gamma=1/2$ is observed for some system ($\tilde{B}_2 >0$), the decrease
of a counterion size would entail the alteration of the sign of $\tilde{B}_2$ and hence disappearance of this
regime, as for $\tilde{B}_2 <0$. On the other hand, if the regime with $\gamma =1/2$ is absent
($\tilde{B}_2<0$), the increase of the counterion size would  lead to the change of the sign of $\tilde{B}_2$
and appearance of the regime with $\gamma =1/2$, as for $\tilde{B}_2>0$.

\begin{figure}
\centering
\includegraphics[width=\columnwidth]{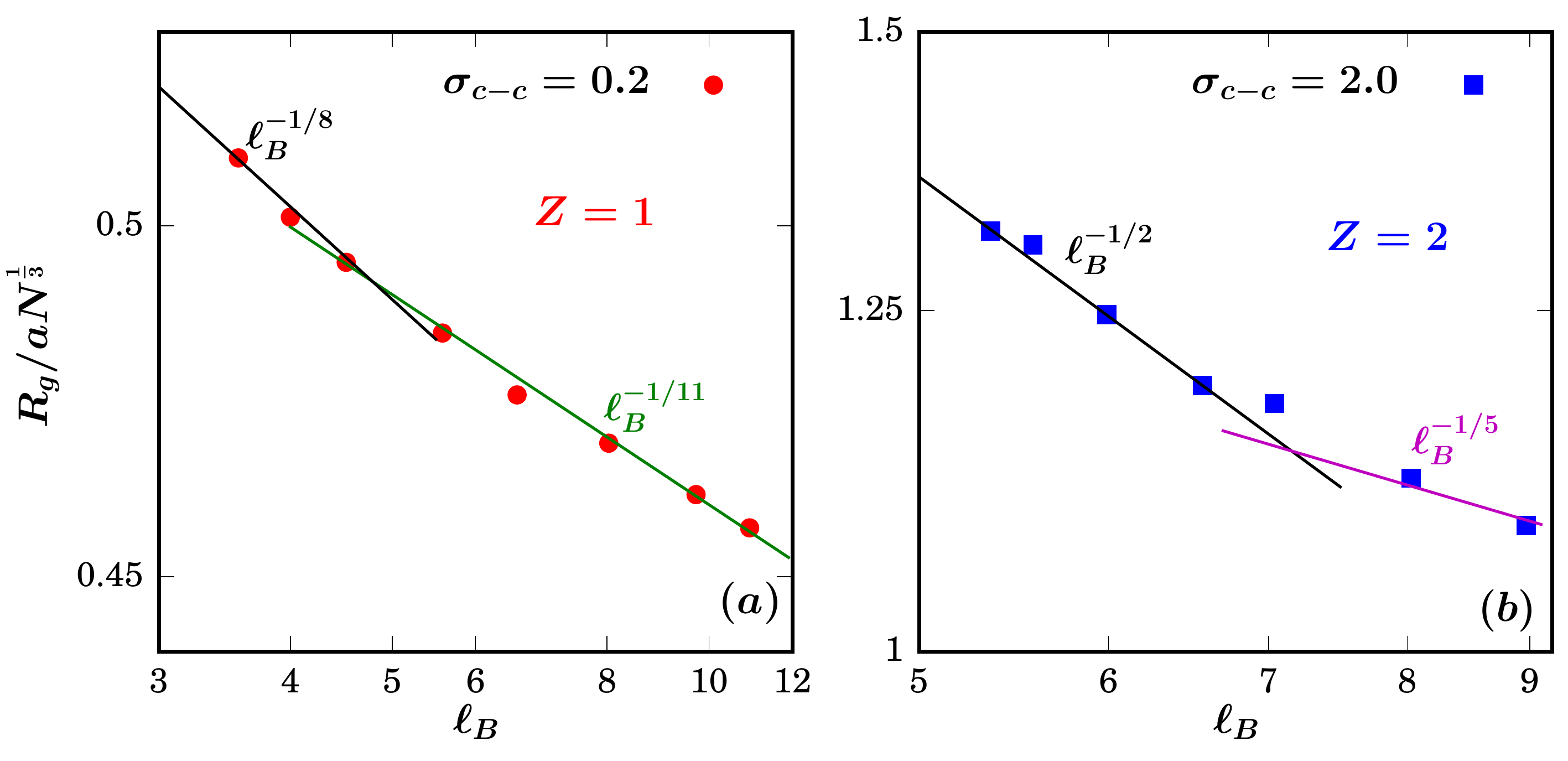}
\caption{(a) The disappearance of $R_g \sim \ell _B^{-1/2}$ and $R_g \sim \ell _B^{-1/5}$ regimes for
monovalent-counterion  system with $\varepsilon_{LJ}=1$ by reducing the counterion radius [compare with
Fig.~\ref{fig:RgZ_Poor}(a)]. (b) The appearance of $R_g \sim \ell _B^{-1/2}$ regime for divalent-counterion
system with $\varepsilon_{LJ}=1.5$ by increasing the counterion radius [compare with
Fig.~\ref{fig:RgZ_Poor}(e)]. The chain length is $N=204$. } \label{fig:LJRg_Poor}
\end{figure}

To confirm these  predictions, we perform additional simulations: Firstly, for the case of monovalent
counterions we decrease the size of counterions  and test whether the regime with $\gamma=1/2$ disappears.
Secondly, for the case of divalent counterions, we increase the size of counterions and check  whether the
regime with $\gamma=1/2$ emerges. The MD data for these two simulations are shown in
Fig.~\ref{fig:LJRg_Poor}. In the case of monovalent counterions, as the size of the counterions is reduced,
we find that the regime with $\gamma=1/2$ vanishes [see Fig.~\ref{fig:LJRg_Poor}(a)], that is, $\tilde{B}_2$
becomes negative, $\tilde{B}_2 <0$. In the case of divalent counterions the regime with $\gamma =1/2$, absent
for $\sigma_c=1$ [see Fig.~\ref{fig:RgZ_Poor}(e)], appears when the counterion size increases up to
$\sigma_c=2$ [see Fig.~\ref{fig:LJRg_Poor}(b)]. These data confirm that the presence of counterions inside
the condensed phase modulates the effective attractive interactions between monomers and can change the sign
of the second virial coefficient $\tilde{B}_2$.

\begin{figure}
\centering
\includegraphics[width=\columnwidth]{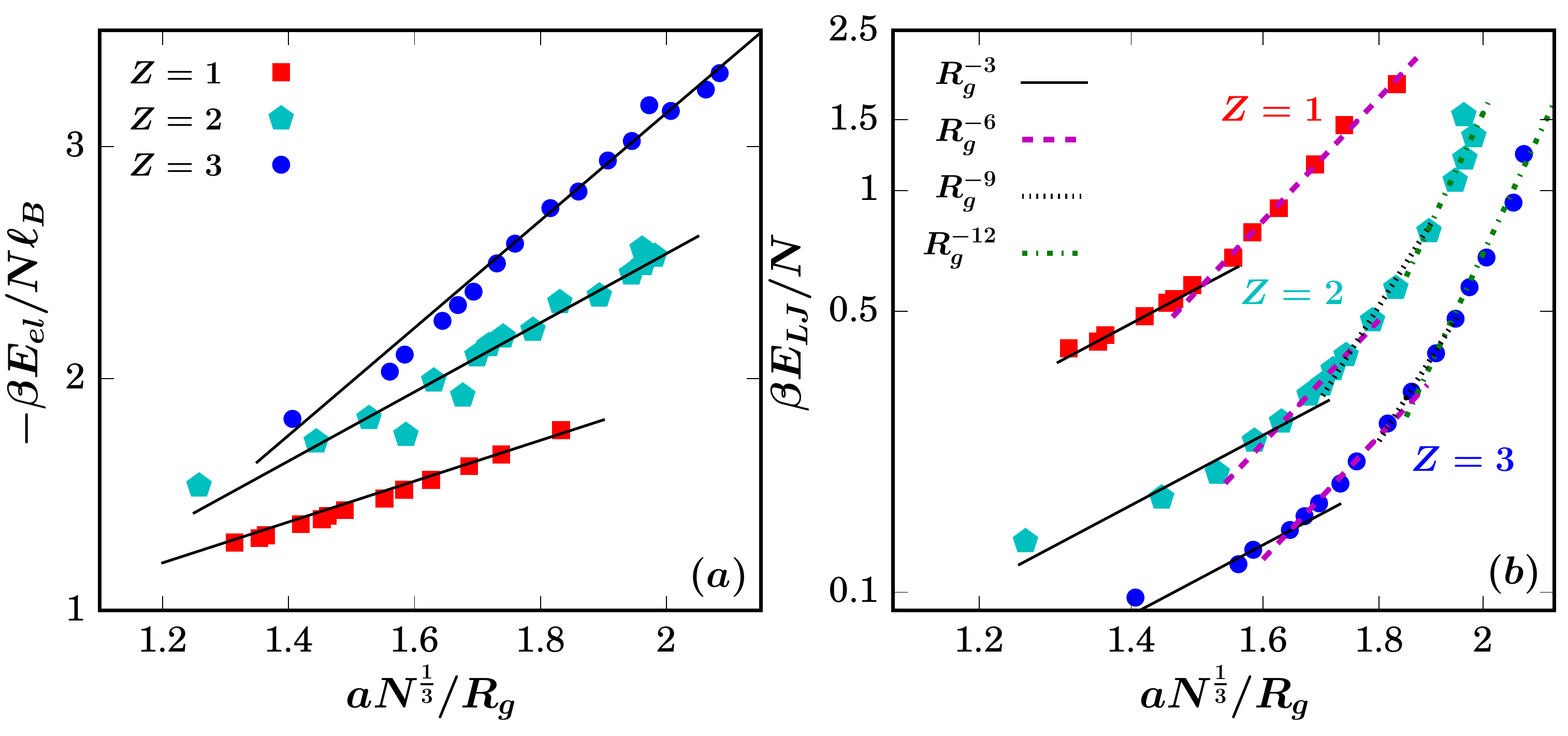}
\caption{The dependence of (a) the electrostatic energy $E_{\rm el}$, (b) LJ energy $E_{LJ}$ of the system on
the gyration radius $R_g$ for different valencies of  counterions. $\epsilon_{LJ}=1.0$ and $r_C=2.5$ for
monomer-monomer interactions. The chain length is $N=204$. } \label{fig:ElRg_Poor}
\end{figure}

The  counterion fluctuation theory, as developed in Sec~\ref{sec:theory}, may be further substantiated by
computing the energies $E_{el}$ and $E_{LJ}$ from MD simulations. As can be  seen from Eq.~\eqref{eq:Eel},
the scaling of $E_{el}$ is independent of range of $\ell_B$ and solvent quality and scales as inverse of
$R_g$. However, the dependence of $E_{LJ}$ on $R_g$ is more complicated  ($E_{LJ} \sim R_g^{-3k}$) and is a
function of dominant $k$-th virial term; this, in turn, depends on the range of $\ell_B$, as can be seen from
Eq.~\eqref{eq:ELJ}. The results for the scaling of $E_{el}$ and $E_{LJ}$ with $R_g$ from our MD simulations
 are shown in Fig.~\ref{fig:ElRg_Poor}.
The data captures both the linear dependence of $E_{el}$ on $R_g$ and dependence of $E_{LJ}$ on various
powers of $R_g$ very well, validating the free energy expression, Eq~\eqref{eq:ftot},  obtained in the
counterion fluctuation theory~\cite{brilliantov98}. For $E_{LJ}$, we also note that as the valency of the
counterions is increased, powers of $R_g$, corresponding to larger virial terms, appear. A possible physical
explanation for this is that as the electrostatic interactions in the system become stronger and the packing
fraction of all species inside the globule increases, more terms to account for the  volume interactions in
the collapsed state are needed.

\section{Discussion and conclusions \label{sec:discussion}}

In this paper we studied theoretically and by means of MD simulations the nature of a collapsed state of a
polyelectrolyte (PE) in a poor solvent for different strength of electrostatic and volume interactions. We
detect several sub-regimes of the collapsed state of the PE, characterized by the scaling relation, $R_g \sim
\ell_B^{-\gamma}$, for the gyration radius, $R_g$, and the reduced Bjerrum length, $\ell_B$.  From the MD
simulations, we find that for different intervals of $\ell_B$ the exponent $\gamma$ takes a series of values
as a function of the solvent quality and the valency of counterions. This is consistent with the predictions
of our theoretical analysis shown in Table~\ref{table1}. In particular, the exponent $\gamma$ has a general
form $\gamma = 1/(3k-4)$, if in the part of the free energy, associated with the volume interactions, the
$k$-th virial term dominates.

In our earlier work~\cite{CollGood} on collapsed regimes of a PE in good solvent, we had shown that
modification of counterion-fluctuation theory~\cite{brilliantov98} can explain the scaling of $R_g$ on
$\ell_B$ observed  in MD simulations. In the case of good solvent,  for the values of $\ell_B$ studied, we
detected only two sub-regimes: weak ($\gamma=1/2$) and strong ($\gamma=1/5$) electrostatic collapse, and the
values of exponent were found to be independent on the properties of the solvent and valency of the
counterions. We explained the existence of these two regimes, modifying the existing counterion-fluctuation
theory  by inclusion of the third virial term into the volume part of the free energy of a PE. All volume
interactions in this work  were repulsive to emulate a good solvent; moreover all volume interactions were
the same for monomers and counterions.  In the current study, we further extend this formalism by explicitly
considering the interactions between monomer-monomers, monomers-counterions and counterions-counterions and
develop a generalized theory for a collapsed regime of a PE in any solvent. The good solvent case considered
earlier is a special case of this generalized counterion-fluctuation theory.

We note that, while in MD simulations the solvent quality can be controlled by the interaction potential
between monomers, in a theory this property is characterized by the sign of the second virial coefficient:
$B_2$ is positive ($B_2>0$) for a good solvent and negative ($B_2<0$) for a poor one.  It is not clear,
however, whether this definition of solvent quality, based only on monomer-monomer interactions, remains
meaningful for charged polymers in the presence of counterions. Within our generalized theory, we expect that
the sign of the renormalized $\tilde{B}_2$ will be manifested in  MD simulations through the  presence  of
the sub-regime, $R_g \sim \ell_B^{-\gamma}$,  with the exponent $\gamma=1/2$ for $\tilde{B}_2 > 0$ and
 absence, respectively,  of this regime for  $\tilde{B}_2 < 0$.

\begin{figure}
\centering
\includegraphics[width=0.7\columnwidth]{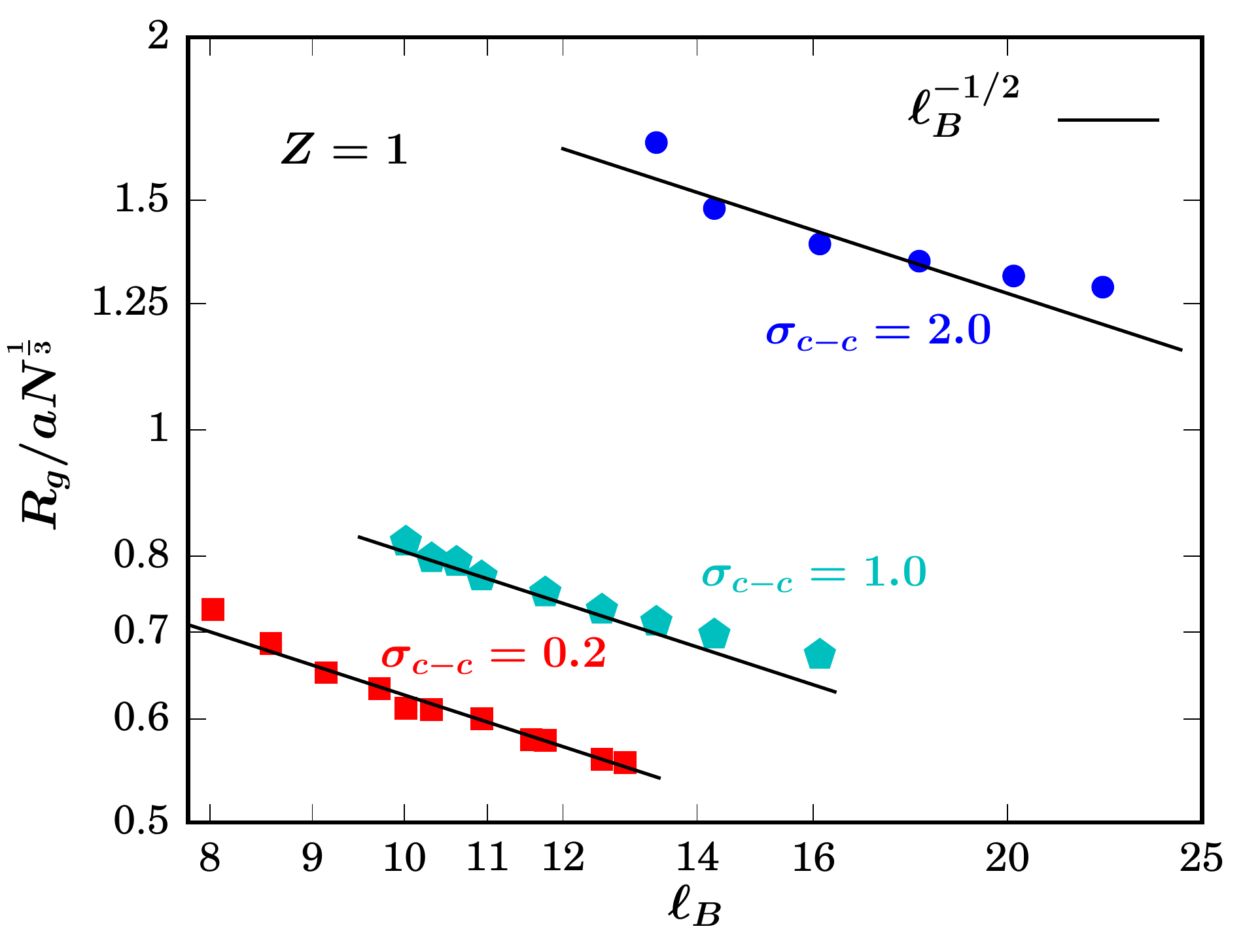}
\caption{The dependence of the gyration radius $R_g$ for different radius of  counterions for good solvent ($\epsilon _{LJ}=1$ and $ r_c=1$). The chain length is $N=204$ and the valency of counterions $Z=1$}
\label{fig:LJRg_Good}
\end{figure}

In the current study, through extensive MD simulations, we demonstrate that the  condensation of counterions
on a PE chain leads to an effective renormalization of the volume virial coefficients. The renormalized
virial coefficients strongly depend on the valency of counterions and the "strength" of the poor solvent,
which may be characterized by the value of $\varepsilon_{LJ}$ -- the energy parameter of the LJ potential.
Surprisingly, the MD results for a particular set of parameters for the volume interaction potential show,
via the presence of sub-regime with the exponent $\gamma=1/2$, that the renormalized second virial
coefficient $\tilde{B}_2$ is positive. This is opposite to the expectation for this coefficient to be
negative, as the simulations were performed for a poor solvent [see Fig.~\ref{fig:RgZ_Poor} (a)]. When the
"strength" of the poor solvent $\varepsilon_{LJ}$ increases, this sub-regime disappears for divalent and
trivalent counterions but persists for monovalent counterions [see Fig.~\ref{fig:RgZ_Poor} (b)]. To
understand the role of condensed counterions on the sign of renormalized $\tilde{B}_2$, we performed a
theoretical analysis as well as additional simulations in which we varied the size of counterions and
demonstrated that the appearence and disappearance of sub-regime with $\gamma=1/2$ crucially depends on the
size of the counterions [see Fig.~\ref{fig:LJRg_Poor}]. This dependence of the sign of the renormalized
$\tilde{B}_2$ on counterion size and valency occurs only for a poor solvent, as the condensed counterions can
modulate the effective attractive interactions between monomers resulting in the alteration of the sign of
$\tilde{B}_2$. For a good solvent, with repulsive interactions between the monomers, the counterion size and
valency play no role, which may be clearly seen from  Fig.~\ref{fig:LJRg_Good}, where we present the
according results of MD simulations. The results in Fig.~\ref{fig:LJRg_Good} combined with those in
Fig.~\ref{fig:LJRg_Poor}, show that while the sign of $\tilde{B}_2$ is unambiguous in a good solvent, the
same is not true in the case of a poor solvent and depends on several system parameters such as strength of
the solvent, valency and size of the counterions. This suggests that for charged polymers with attractive
monomer-monomer volume interactions and in the presence of counterions, the sign of the second virial
coefficient cannot be assumed to be always negative. This is in a striking contrast with collapsed neutral
polymers, which have the same attractive monomer-monomer interactions,  where the sign of $B_2$ is always
negative.

We also validate the predictions of the generalized counterion-fluctuation theory,  developed in this paper,
through the comparison of the theoretical and MD simulation results for scaling of different parts of the
internal energy of the system with $R_g$. The original counterion-fluctuation theory~\cite{brilliantov98}
predicts that the electrostatic internal energy of the system $E_{el}$ scales with the gyration radius as
$E_{el}  \sim R_g^{-1}$ regardless of the sub-regime. At the same time, the dependence of internal energy
associated with the volume (LJ) interactions is expected to be different for different sub-regimes, similar
to the dependence of $R_g$ on $\ell_B$ described above. From the theoretical analysis in we find that the LJ
energy is expected to scale with $R_g$ as $E_{LJ} (R_g) \sim R_g^{-3k}$, where $k$ refers to the dominant
$k$-th virial term in the part of the free energy that refers to the volume interactions. The MD simulation
results  are in complete agreement with these predictions [see Fig.~\ref{fig:ElRg_Poor}]. We also note that
the values of $R_g$ at which the crossovers from one sub-regime  of $E_{LJ}(R_g)$ to another take place [see
 Fig.~\ref{fig:ElRg_Poor}] coincide with the values of $R_g$ where the crossovers between regimes with
different exponents $\gamma$ are  detected [see Fig.~\ref{fig:RgZ_Poor}].

Based on our findings, we conclude that the effective attractive electrostatic interactions in systems of
like-charged polymers in the presence of counterions is described well in terms of correlated fluctuations of
counterions, as has been proposed in the counterion fluctuation theory~\cite{brilliantov98}. The
electrostatic term of the free energy of the system, based on the counterion fluctuation, is independent of
the solvent quality. We note  that none of the other existing theories of effective electrostatic
interactions of a PE~\cite{pincus98,delaCruz2000,Cherstvy2010,muthukumar2004,arti14} can explain the sequence
of electrostatic sub-regimes or the scaling of the electrostatic energy with $\ell_B$, as seen in our  MD
simulations.

%\emph{Acknowledgments.}
\section{Acknowledgments.}
%\begin{acknowledgments}
The simulations were carried out on the supercomputing machines Annapurna, Nandadevi and Satpura at the Institute of Mathematical Sciences.
%\end{acknowledgments}

\section{Appendix}
\label{App:I}

The free energy for the volume interactions among monomers may be written in the form of virial expansion as,
\bea \lbl{eq:Fvol_mm} \beta F_{\rm vol\, m.m.} &=& \sum_{k=2}^{\infty}  B_k \rho_m^k \, V_{\rm g}
= \sum_{k=2}^{\infty}  B_k \left(\frac{N}{V_{\rm g}} \right)^k \, V_{\rm g} \nonumber \\
&=&\sum_{k=2}^{\infty} \left(\frac{4 \pi a^3}{ 3 \sqrt{6}} \right)^{1-k }
 \frac{N^{(3-k)/2}}{\alpha^{3(k-1)}}B_k ,
\eea where $B_k$ is the $k$-th virial coefficient for  monomer-monomer interactions and  $\rho_m = N/V_{\rm
g}$ is the average density of monomers inside the gyration volume. Similar to Eq.~\eqref{eq:Fvol_mm}, the
free energy of the volume interactions of counterions reads,
\bea
\lbl{eq:Fvol_cc}
\beta F_{\rm vol\, c.c.}
&=& \sum_{k=2}^{\infty}  C_k \rho_{\rm c. in}^k \, V_{\rm g}
= \sum_{k=2}^{\infty}   C_k \left(\frac{N_c}{V_{\rm g}} \right)^k \, V_{\rm g} \nonumber \\
&=&\sum_{k=2}^{\infty} \left(\frac{4 \pi a^3}{ 3 \sqrt{6}} \right)^{1-k } \frac{C_k}{Z^k}
\frac{N^{(3-k)/2}}{\alpha^{3(k-1)}} ,
\eea
where $\rho_{\rm c. in} \simeq N_c/V_{\rm g}= N/ZV_{\rm g}$ is the average counterion density inside the
gyration volume and we approximate it by the according density, when all counterions are condensed. $C_k$ are
the virial coefficients for the counterion-counterion interactions. Furthermore, the monomer-counterion
volume interactions are described by the term, \bea \lbl{eq:Fvol_mc}
\beta F_{\rm vol\, c.m.} &=& \sum_{k=2}^{\infty} \sum_{l=0}^{k} {\cal C}_l^k \rho_c^l \rho_m^{k-l} D_{l, k-l} V_{\rm g} \\
&=& \sum_{k=2}^{\infty}  \left(\frac{4 \pi a^3}{ 3 \sqrt{6}} \right)^{1-k }
\frac{N^{(3-k)/2}}{\alpha^{3(k-1)}} \sum_{l=0}^{k} \frac{{\cal C}_l^k}{Z^l}D_{l, k-l}, \nonumber
\eea
where ${\cal C}_l^k = k!/l!(k-l)!$ are the combinatorial coefficients  and $D_{l, k-l}$ is the $k$-th virial
coefficient for monomer-counterion volume interactions which refers to $l$ counterions and $k-l$ monomers.

Using Eqs.~(\ref{eq:Fvol_mm}), (\ref{eq:Fvol_cc}) and (\ref{eq:Fvol_mc}) one can write the part of the free
energy responsible for the volume interactions in the system in the following compact form,
\bea
\lbl{eq:Fvol_all11}
\beta F_{\rm vol} = \sum_{k=2}^{\infty} \frac{N^{(3-k)/2}}{\alpha^{3(k-1)}} \,
\tilde{B}_k ,
\eea
where the renormalized virial coefficients $\tilde{B}_k$, that account for  all volume
interactions, are defined as
\be
\lbl{eq:Btild11} \tilde{B}_k =\left(\frac{2 \pi a^3}{ 9 \sqrt{6}}
\right)^{1-k } \left( B_k + \sum_{l=0}^{k} \frac{{\cal C}_l^k}{Z^l}D_{l, k-l} + \frac{C_k}{Z^k} \right).
\ee

\begin{figure}
\includegraphics[width=0.95\columnwidth]{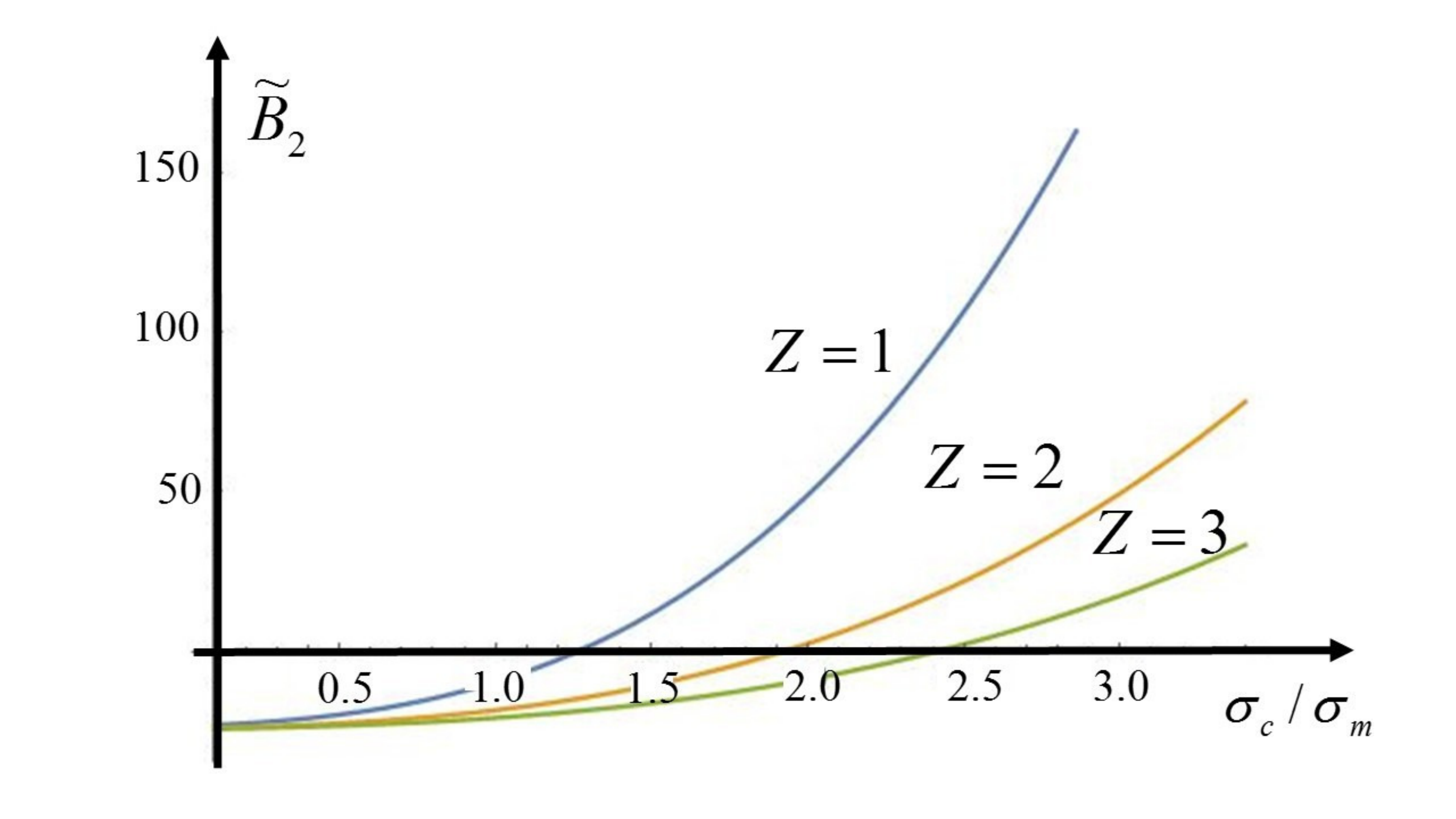}
\caption{The dependence of the renormalized second virial coefficient $\tilde{B}_2$ on the ratio  of the
counterion and monomer sizes, $\sigma_c/\sigma_m$ and the counterions valency $Z$, as it follows from
Eq.~\eqref{eq:Btild1}. The bare coefficients $B_2$, $C_2$ and $D_{1,1}$ are computed using
Eq.~\eqref{eq:B2def} with the according Lennard-Jones potentials: $\varepsilon_{LJ} =1.5$ and $r_c=2.5$ for
the monomer-monomer interactions, $\varepsilon_{LJ}=1 $ and $r_c=1  $ for the counterion-counterion
interactions and $\varepsilon_{LJ} = \sqrt{1\cdot 1.5}$ and $r_c=(2.5+1)/2=1.75$ for the monomer-counteiron
interactions. Depending on the valency of the counterions the renormalized coefficient $\tilde{B}_2$ changes
its sign with the increasing size of the counterions. } \label{fig:B2B3}
\end{figure}

Here we consider a collapsed state of a PE chain and the main difference of a globular  state, as compared to
that of a coiled state is that a  "globule contains a number of uncorrelated parts of the chain" so that a
concept of quasimonomers~\cite{Khokhlov1977} and the according virial expansion for pressure or free energy
is   valid~\cite{GrosbergKuznetsov_MM92}.  In our study we use the LJ potential \eqref{eq:LJpot} to model all
volume interactions. For the monomer-monomer interactions we use  the cutoff distance $r_c= 2.5 \sigma$, that
is,  these volume interactions include both attractive and repulsive forces. For the monomer-counterion and
counterion-counterion interactions we use the cutoff $r_c=  \sigma$, which corresponds to purely repulsive
forces.  The virial coefficient $B_2$ reads\footnote{For simplicity we ignore the contribution to the second
virial coefficient $B_2$ from the third virial coefficient $B_3$ that appears due to the chain
connectivity~\cite{GrosbergKuznetsov_MM92}. It may be shown that for the addressed range of parameters this
contribution is positive and not large to yield a qualitative difference.} \bea \lbl{eq:B2def} B_2 &=&
-\frac12 \int d{\bf r} f(r)=- \int d{\bf r} \left(e^{-\beta U_{LJ}(r)} -1\right) \nn
 &=&-2\pi \int_0^{r_c} dr r^2 \left(e^{-\beta U_{LJ}(r)} -1\right).
\eea
Similar expressions apply for the coefficients $C_2$ and $D_{1,1}$, with the according  change of the LJ
potential. Using these expressions for the virial coefficients $B_2$, $C_2$ and $D_{1,1}$ one can compute,
using Eq.~\eqref{eq:Btild1} the renormalized second virial coefficient $\tilde{B}_2$. The results are
presented in Fig.~\ref{fig:B2B3}, where we demonstrate the dependence of $\tilde{B}_2$ on the size of the
counterions $\sigma_c$ and their valency $Z$. As it may be seen from the Fig.~\ref{fig:B2B3} the renornalized
coefficient $\tilde{B}_2$ changes its sign from negative to positive with increasing size of counterions.
This effect corresponds to an effective change of the solvent quality due to counterion condensation.

%%%END OF MAIN TEXT%%%


%\balance


%%%REFERENCES%%%
\bibliography{ref} %You need to replace "rsc" on this line with the name of your .bib file
\bibliographystyle{rsc} %the RSC's .bst file

\end{document}